\documentclass[reviewcopy]{elsart}

\usepackage{amssymb}
\usepackage{amsmath}
\usepackage{graphics}

\usepackage{epsfig}
\usepackage{amssymb}
\input epsf

\begin{document}

\begin{frontmatter}



\title{ Proton--induced deuteron breakup at GeV energies
        with forward emission of a fast proton pair\thanksref{sasha}}
\author[dubna]{V.~Komarov\corauthref{komarov}},
\corauth[komarov]{Corresponding author: V. Komarov}
\ead{v.komarov@fz-juelich.de}
\author[ikp,dubna]{S.~Dymov},
\author[erlangen,tbilisi]{A.~Kacharava},
\author[dubna]{A.~Kulikov},
\author[dubna,tbilisi]{G.~Macharashvili},
\author[dubna]{A.~Petrus},
\thanks[sasha]{This paper is dedicated to A.~Petrus who was killed 
 in a tragic accident on May 19, 2002.}
\author[ikp]{F.~Rathmann},
\author[ikp]{H.~Seyfarth},
\author[ikp]{H.~Str\"oher},
\author[dubna,almata]{Yu.~Uzikov},
\author[erlangen,dubna]{S.~Yaschenko},
\author[dubna]{B.~Zalikhanov},
\author[ikp]{M.~B\"uscher},
\author[zam]{W.~Erven},
\author[ikp]{M.~Hartmann},
\author[munster]{A.~Khoukaz},
\author[ikp]{R.~Koch},
\author[dubna]{V.~Kurbatov},
\author[munster]{N.~Lang},
\author[ikp]{R.~Maier},
\author[dubna]{S.~Merzliakov},
\author[gatchina]{S.~Mikirtytchiants},
\author[rossendorf]{H.~M\"uller},
\author[tbilisi]{M.~Nioradze},
\author[ikp]{H.~Ohm},
\author[ikp]{D.~Prasuhn},
\author[munster]{R.~Santo},
\author[koln]{H.~Paetz gen.~Schieck},
\author[ikp]{R.~Schleichert},
\author[ikp]{H.J.~Stein},
\author[ikp]{K.~Watzlawik},
\author[dubna]{N.~Zhuravlev}, and
\author[zam]{K.~Zwoll}

\address[dubna]{Joint Institute for Nuclear Research, LNP, 141980 Dubna, Russia}
\address[ikp]{Institut f\"ur Kernphysik, FZJ, 52425 J\"ulich, Germany}
\address[erlangen]{Phys. Inst. II, Universit\"at Erlangen--N\"urnberg, 91058 Erlangen, Germany}
\address[tbilisi]{High Energy Physics Institute, Tbilisi State University, 380086 Tbilisi, Georgia}
\address[almata]{Kazakh National University, 480078 Almaty, Kazakhstan}
\address[zam]{Zentrallabor f\"ur Elektronik, FZJ, 52425 J\"ulich, Germany}
\address[munster]{Institut f\"ur Kernphysik, Universit\"at M\"unster, 48149 M\"unster, Germany}
\address[gatchina]{St. Petersburg Nuclear Physics Institute, 188350 Gatchina, Russia}
\address[rossendorf]{Institut f\"ur Kern-- und Hadronenphysik, FZR, 01474 Dresden, Germany}
\address[koln]{Institut f\"ur Kernphysik,Universit\"at zu K\"oln, 50937  K\"oln, Germany}

\begin{abstract}
 A study of the deuteron breakup reaction
 $pd \to (pp)n$   with forward emission of a fast proton pair 
 with small excitation energy $E_{pp}<$ 3~MeV
 has been performed using the  ANKE  spectrometer at COSY--J\"ulich.
 An exclusive measurement was carried out
 at six proton--beam energies $T_p=$~0.6,~0.7,~0.8,~0.95,~1.35, and 1.9~GeV
 by reconstructing the momenta of the two protons.
 The differential cross section of the breakup reaction, 
 averaged up to $8^{\circ}$ over the cm polar angle  
 of the total momentum of the $pp$ pairs, has been obtained.
 Since the kinematics of this process is quite similar to that of backward 
 elastic $pd \to dp$ scattering, the results are compared to 
 calculations based on a theoretical model previously 
 applied to the $pd \to dp$ process.
\end{abstract}


\begin{keyword}
Deuteron breakup;  Short--range nucleon--nucleon interaction

\begin{PACS}
13.75.Cs, 25.10.+s, 25.40-h.\\[1ex]
\end{PACS}
\end{keyword}
\end{frontmatter}


\section{Introduction}
\begin{figure}[htb]
\begin{center}
\epsfxsize=\columnwidth \epsfbox{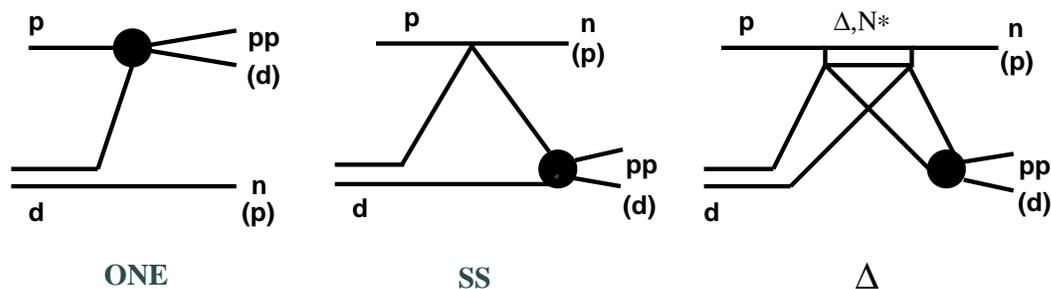}
\caption{\label{fig1} Mechanisms included in the ONE+SS+$\Delta$ model
for the $pd \to (pp)n$ ($pd \to dp$) processes.}
\end{center}
\end{figure}

 Backward elastic $pd \to dp$ scattering at energies of several hundred
 MeV is one of the simplest hadron--nucleus processes with high
 transferred momentum.  It has been studied for more than 30 years
 both experimentally and theoretically with the aim of extracting
 information about the short--range structure of the $NN$ interaction
 and the dynamics of high--momentum transfer in few--nucleon systems.
 Besides the one--nucleon--exchange (ONE) mechanism (Fig.~\ref{fig1}),
 a number of concepts have been discussed in this context, e.g. the presence
 of nucleon resonances ($N^*$) inside the deuteron \cite{kermankiss}, 
 the importance of virtual pions \cite{cwilkin69}, and three--baryon
 resonances \cite{kondrat} (for a review see Ref. \cite{uzikov}).
 Only at low energies, where ONE
 dominates, are the data on differential cross
 section, tensor analyzing power $T_{20}$,
 and spin transfer coefficient $\kappa$,
 reasonably well described [4-8].  
 At higher energies, where internal momenta above 0.3~GeV/c
 are probed in the deuteron, the dynamics becomes more
 complicated, because of a possible excitation of $N^*$ and $\Delta$
 resonances in the intermediate states. 
 These effects are taken into
 account to some extent in the one--pion--exchange model, 
 but when adding the ONE amplitude, the problem of double 
 counting arises \cite{cwilkin69,kolib,naka}.
 The excitation of the $\Delta(1232)$ resonance in the intermediate state
 ($\Delta $ mechanism) is explicitly included in a model
 \cite{kondrat,boud}, which also takes into account coherently ONE and
 single $pN$ scattering (SS) in a consistent way (Fig.~\ref{fig1}).  
 This model, improved in Ref. \cite{imam11} 
 with respect to the $\Delta$ contribution through the
 analysis of $pp\to pn\pi^+$ data \cite{imam1}, describes  
 the gross features of the $pd \to dp$ 
 spin--averaged differential cross section.
 After further refinement also the tensor analyzing power
 at beam energies below 0.5~GeV is qualitatively reproduced \cite{boud}.  
 Above the region, where the $\Delta$(1232) dominates, the role of intermediate 
 excitations of heavier baryon resonances is expected to increase and
 this makes the theoretical interpretation of this process much more ambiguous.\\  
 In view of the above complications,
 it would be very important to study
 a similar $pd$ process, where contributions from the $N^*$ 
 and $\Delta $ resonance excitation are suppressed.
 For that purpose, an appropriate reaction is
 the deuteron breakup
 $$ p+d \to (pp)+n $$
 with emission of the two protons in forward direction 
 ($\theta_{pp}\approx0^\circ$) at low excitation energy $E_{pp}<3$~MeV.
 With the neutron emitted backward, the kinematics
 of this reaction is quite close to that of $pd$ backward elastic
 scattering. 
 Therefore, the same mechanisms can be applied in
 the analysis of the process as well.
 According to the ONE+SS+$\Delta$ model calculations \cite{imam2,smuz98},
 which implicitly include the $pp$ final--state interaction (fsi),
 the $pp$ pair is expected to be mainly in a ${}^1S_0$ state.
 Due to isospin invariance, 
 the isovector nature of the $pp$ pair
 leads to a suppression of the amplitude of the $\Delta$
 mechanism by a factor three in  comparison to the ONE amplitude 
 for all partial waves of the $pp$ system \cite{imam2}.
 The same suppression factor also applies for a broad class of diagrams with
 isovector meson--nucleon rescattering in the intermediate state,
 including excitation of $N^*$ resonances \cite{uzzhetf}.
  As a result, the contribution of the ONE mechanism, which is sensitive
 to the $NN$ potential at short distances, becomes more pronounced
 than in $pd \to dp$ scattering.
 Furthermore, 
 the node in the half--off--shell $pp$ scattering amplitude in
 the ${}^1S_0$ state at an off--shell momentum of about 0.4~GeV/c
 leads to a dip of the differential cross section of the deuteron breakup at
 0.7--0.8~GeV beam energy \cite{imam2,uzikov2}.
 At higher energies of 1--3~GeV, the cross section is dominated
 by the ONE mechanism and decreases rather smoothly.\\
 Another attractive feature of the process is the
 simplicity of its phenomenological description, 
 since at zero degrees it requires
 only two spin amplitudes. Therefore, a model--independent
 amplitude analysis becomes possible through the measurement 
 of a few polarization observables. 
 As a first step, 
 we have measured the differential cross section at six
 beam energies in the interval 0.6--1.9~GeV, which covers the region
 of the dip predicted by the ONE+SS+$\Delta$ model, thereby probing a
 wide range of high internal momenta of the $NN$ system ($q_{NN} \sim$ 0.3--0.6~GeV/c).

\section{Experiment}
\begin{figure}[htb]
\begin{center}
\epsfxsize=\columnwidth \epsfbox{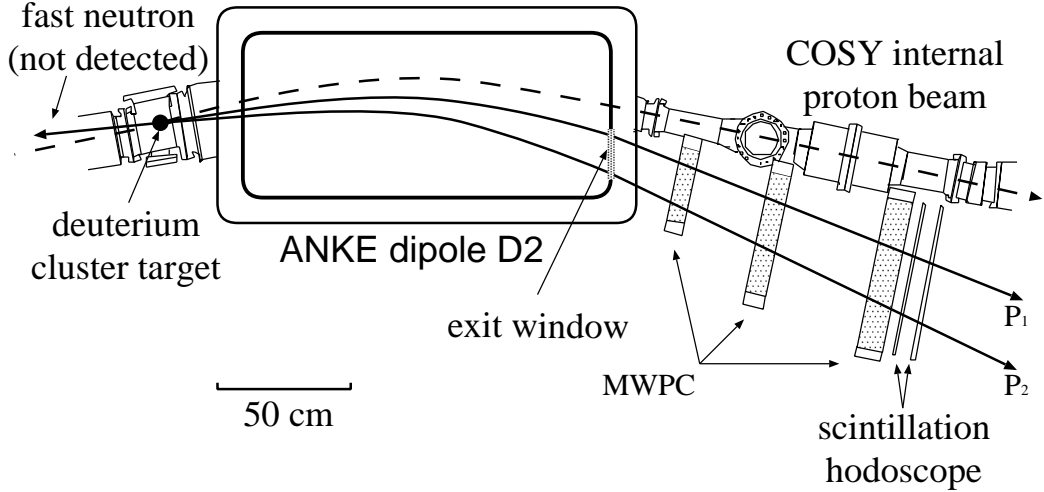}
\caption{\label{fig2} Top view of the experimental setup with the 
 forward detection system of the ANKE spectrometer.}
\end{center}
\end{figure}

\begin{figure}[hbt]
\begin{center}
\epsfxsize=\columnwidth \epsfbox{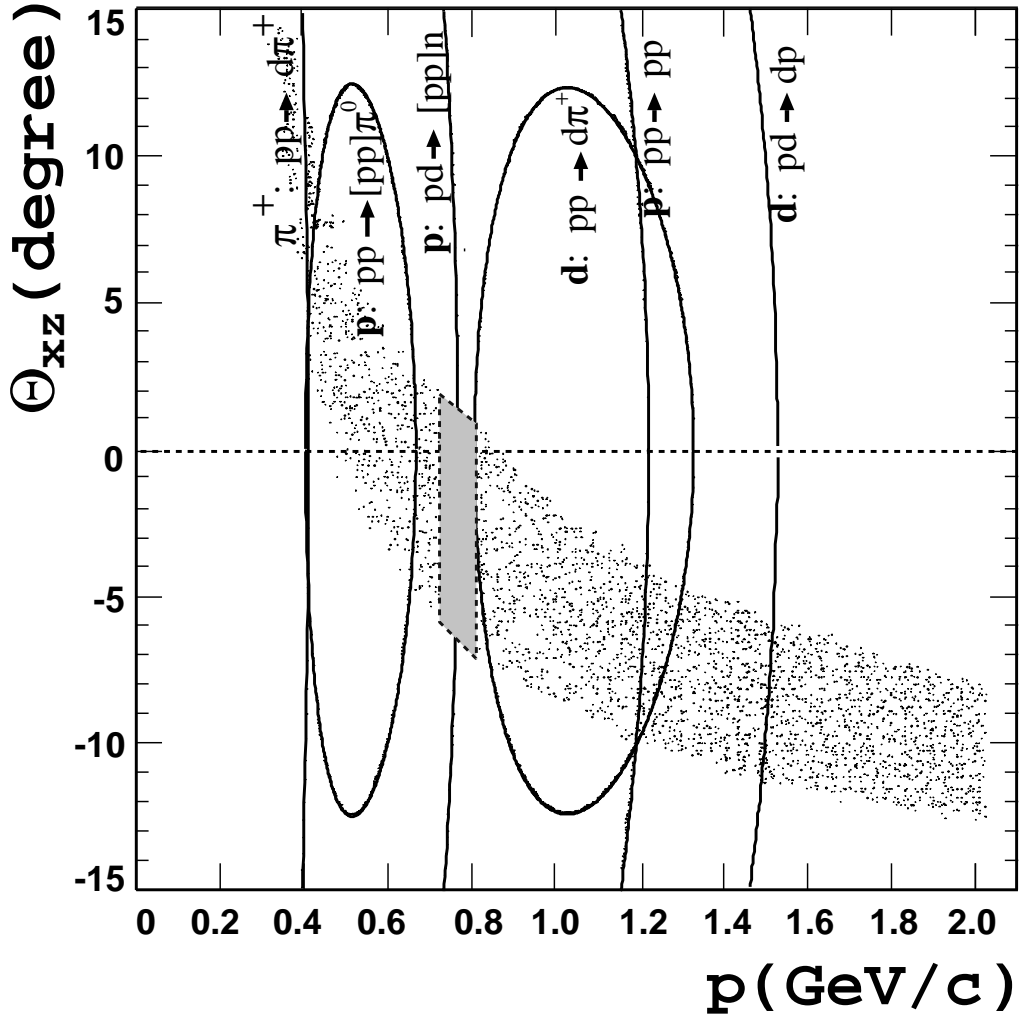}
\caption{\label{fig3} Plot of the acceptance of the setup 
 from a MC simulation showing polar angle versus momentum
 at $0.6\,$ GeV beam energy.
 $\Theta_{xz}$ is the scattering angle of 
 the emitted particle projected onto the 
 median plane of the spectrometer. 
 The curves show kinematical loci for 
 $\pi^{+}$, $\rm p$, and $\rm d$
 from the indicated processes. 
 The symbol [pp] denotes $pp$ pairs with zero excitation energy, 
 while the grey area contains those of $E_{pp}<3$~MeV.}
\end{center}
\end{figure}

\begin{figure}[htb]
\begin{center}
\epsfxsize=\columnwidth \epsfbox{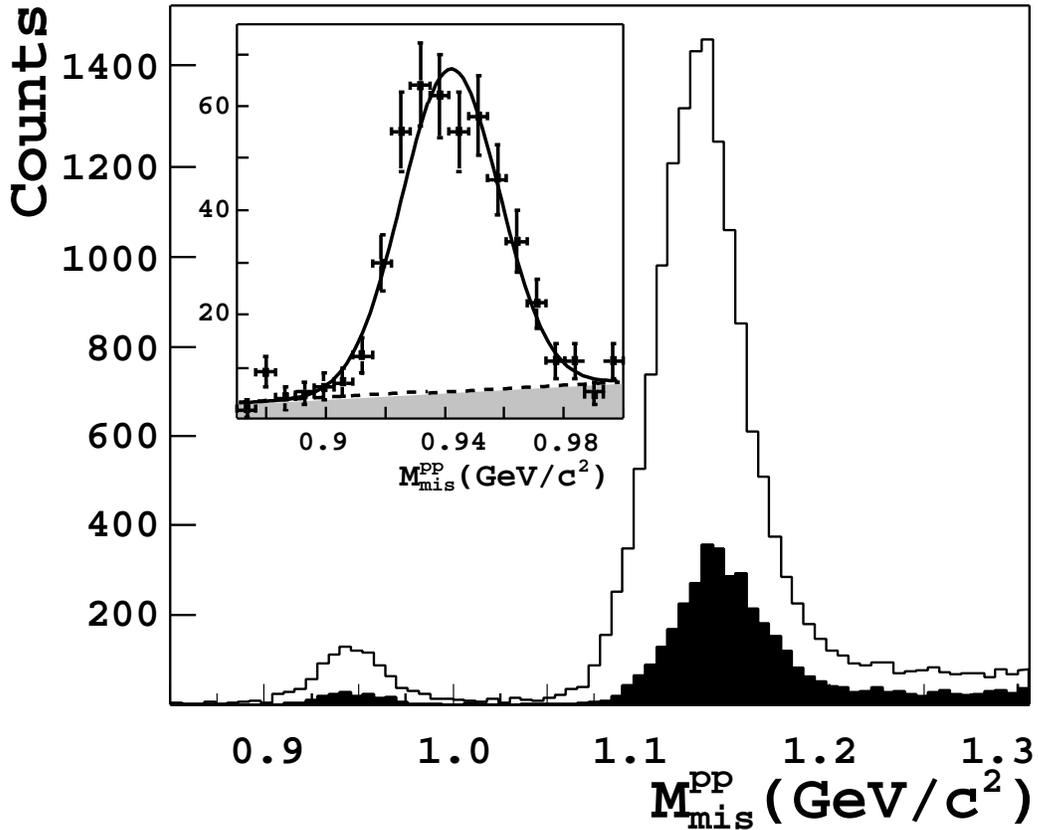}
\caption{\label{fig4} Missing--mass distribution at $T_p = 0.8$~GeV of all 
 identified proton pairs (unfilled histogram).  
 The black histogram denotes identified $pp$ pairs 
 with excitation energy of less than $3$~MeV.
 The inset
 shows the distribution near the neutron mass without particle
 identification for pairs with $E_{pp}<3$~MeV. 
 The background contribution is shown in grey.}
\end{center}
\end{figure}

 The experiment was performed at incident proton beam energies of 0.6,
 0.7, 0.8, 0.95, 1.35, and 1.9~GeV with the spectrometer ANKE
 \cite{barsov} at the internal beam of the
 COoler SYnchrotron COSY--J\"ulich \cite{maier}.  In Fig.~\ref{fig2} those
 parts of the spectrometer are shown that are of concern for the
 present experiment.
 The protons stored in the COSY ring ($\sim 3\cdot10^{10}$)
 impinged on a deuterium cluster--jet target \cite{khou},  
 which provided a target thickness 
 of about $1.3\cdot10^{13}$~atoms/cm$^2$.
 The produced charged particles, after passing the 
 magnetic field of the dipole D2, were registered by a set of three multiwire
 proportional chambers (MWPC) and a scintillation--counter hodoscope. 
 Each wire chamber contains a horizontal and a vertical anode--wire
 plane (1 mm wire spacing), and two planes of inclined strips,
 that allowed us to obtain the required resolution 
 of $\approx$ 0.8--1.2\% (rms) in the momentum range 0.6--2.7~GeV/c. \\
 The hodoscope consists of two layers, containing 8 and 9 
 vertically oriented scintillators (4 to 8~cm width, 1.5 to 2~cm thickness).
 It provided a trigger signal, an energy loss measurement, 
 and allowed for the determination of the differences in 
 arrival times for particle pairs hitting different counters.
 Off--line processing of the amplitude data permitted the measurement 
 of the energy--loss with an accuracy of 10 to 20\% (FWHM),
 and of the time--of--flight difference of events
 with two registered particles with a precision of 0.5~ns (rms). 
 A separate measurement with a hydrogen target
 at beam energies of 0.5 and 2.65~GeV was carried out  
 to calibrate the energy loss in the  counters and the momentum scale
 via the processes  $pp\to pp$, $pp\to d\pi^+$, and  $pp\to pn\pi^+$.\\
 The horizontal acceptance of the setup is shown in Fig.~\ref{fig3}.
 The vertical acceptance  corresponds to $\pm3.5^{\circ}$.
 The trigger rate resulted mainly from  
 elastically and quasi--elastically  scattered protons, 
 from protons associated with meson production and, at beam energies 
 below 1 GeV, from deuterons produced in the $pp \to d\pi^+$ reaction.
 Events with two registered particles 
 contributed little to the total trigger rate and were selected off--line.
 Protons from the breakup process $pd \to ppn$ with
 an excitation energy $E_{pp}<3$~MeV could be detected  
 with the experimental setup
 for laboratory polar angles between 0 and $7^{\circ}$ at all energies.\\
 Among those events with two registered particles,
 breakup events are identified by the determination of the missing--mass
 value, calculated under the assumption that these particles are protons. 
 At all energies the missing--mass
 spectra  reveal a well defined peak at the neutron mass
 with an rms value of about 20~MeV (Fig.~\ref{fig4}). 
 The peak is clearly separated from the one
 at 1.1--1.2~GeV/c$^2$, caused by proton pairs from
 the $pd\rightarrow pp\pi^0n$ or $pd\rightarrow pp\pi^-p$ reactions.
 A direct identification of the particle type
 is possible for those events for which the two particles 
 hit different counters in the hodoscope.
 These amount to about 60\%
 of all events in the peak at the neutron mass.
 For $E_{pp}<3$~MeV, the fraction
 varies from 60 to 22\% for $T_p$ = 0.6 to 1.9~GeV.
 The time--of--flight
 difference $\Delta t$ measured in the hodoscope was compared
 to the difference $\Delta t(p_{1},p_{2})$ obtained 
 from the reconstructed particle momenta $p_1$ and $p_2$, 
 again assuming that the two particles are protons.
 Applying a 2$\sigma$ cut to the peak of 
 the $\Delta t - \Delta t(p_{1},p_{2})$
 distribution, proton pairs could be selected such that the contribution
 from other pairs was less than 1\%.
 When both tracks hit the 
 same counter, the energy loss distributions were analyzed and
 found to be in agreement with   
 the assumption that both registered particles were protons.
 However, the energy loss cut was not used,
 since the proton separation from other particles was not quite perfect.
 In this case we relied on the fact that misidentified pairs 
 ($p\pi^+$, $d\pi^+$, $dp$ or $^3{\rm H}\pi^+$) 
 show up only at substantially higher missing mass values
 and therefore cannot contribute to the peak at the neutron mass.
 For background subtraction,
 the spectra in the vicinity of the neutron mass 
 were fitted by the sum of a Gaussian and a straight line 
 (see inset in  Fig.~\ref{fig4}).
 The number of proton pairs and the signal--to--background ratio
 $N_{\rm sig}/N_{\rm bg}$ were determined in a $\pm 2 \sigma$ range around
 the neutron mass.
 The distribution of distances between hits by the proton pairs ($E_{pp}<3$~MeV)
 in the MWPC's yields rms values of 4.9 and 3.3~cm, at 0.6 and 
 1.9~GeV beam energies,  respectively.
 Therefore, a significant loss of $pp$ pairs due to the two tracks being too close
 is expected to occur only below $E_{pp}=0.2$~MeV.
 Since a resolution of 0.2~(0.3)~MeV at $E_{pp}=0.5~(3)$~MeV 
 was achieved, proton pairs with $E_{pp}<3\,$~MeV could be reliably selected.\\
 The integrated luminosity $L^{\rm int}$  
 was obtained  by counting protons, elastically and quasi--elastically 
 scattered  at small laboratory angles between 5 and $10^{\circ}$.
 It is not possible to distinguish
 these processes experimentally at ANKE, but 
 the achieved momentum resolution makes possible a clean separation 
 from the meson production continuum. 
 The number of counts obtained was related to a simulation
 using the calculated small angle $pd \to pX$ cross section.
 The calculation takes into account the sum of elastic and inelastic terms
 in closure approximation of the Glauber--Franco theory \cite{franko},
 which includes the sum over the complete set of final $pn$ states.
 In order to estimate the obtained accuracy, the cross sections, calculated
 for elastic and quasielastic $pd$ scattering within the same framework,
 were compared with the experimental data of Refs.
 \cite{bosch,greben,irom,benet,dalkha} and  \cite{alada} 
 respectively, in the appropriate energy and angle range.
 The resulting $\chi ^2/n.d.f.$=0.85 (n.d.f.=64) and
 $\chi ^2/n.d.f.$=0.73 (n.d.f.=8), respectively, yield
 a 7\% uncertainty of the calculated cross sections.
 The total errors of the luminosities of Table 1 take into account 
 this uncertainty and other systematic errors of 5\%, resulting  
 from  a small variation of the derived luminosity with the polar 
 angle, caused by the position--dependent efficiency of the MWPC.

\section{Results and discussion}
\begin{figure}[htb]
\begin{center}
\epsfxsize=\columnwidth \epsfbox{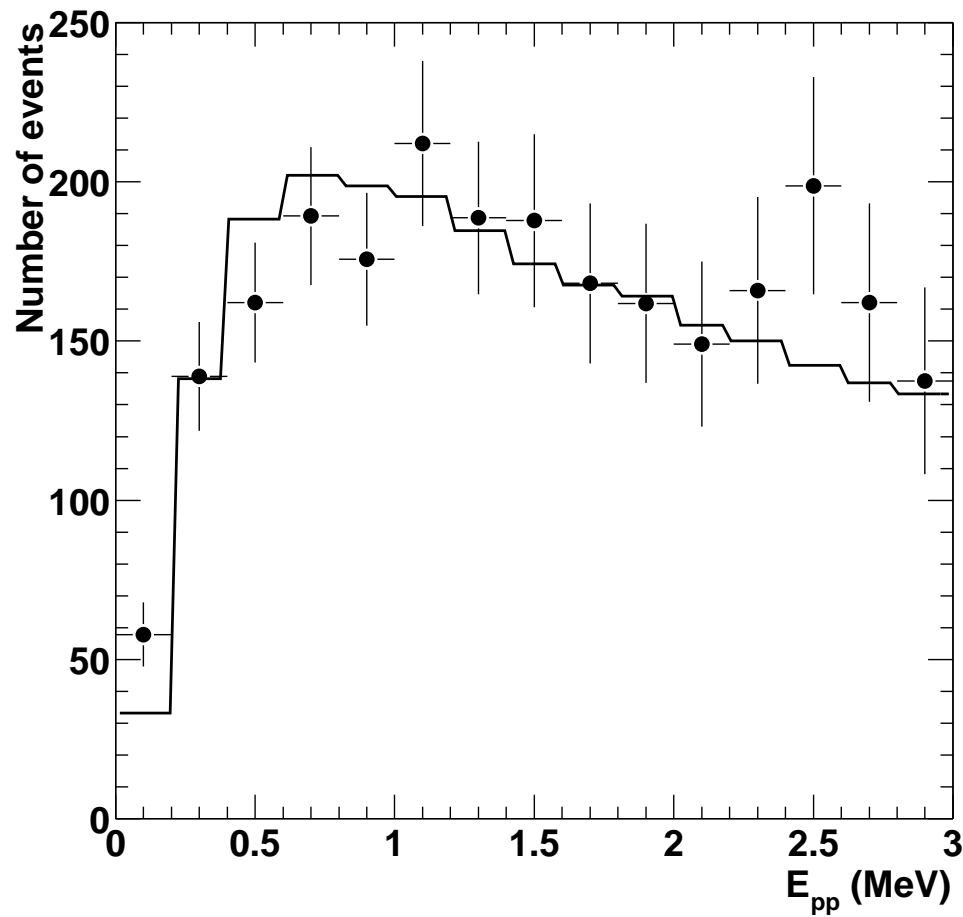}
\caption{\label{fig5} Excitation energy distribution in comparison
 with the theoretical expectation (histogram) from fsi.} 
\end{center}
\end{figure}

 The data allowed us to deduce the three--fold differential cross sections \\
 $d^{3}\sigma/(d \cos \theta_{pp}^{\rm cm} \cdot d\phi_{pp}^{\rm cm} \cdot dE_{pp}$), 
 where $\theta_{pp}^{\rm cm}$ and $\phi_{pp}^{\rm cm}$ are the polar 
 and azimuthal cm angles of the 
 total momentum of the $pp$ pair, respectively.
 (The neutron emission angles correspond  to  $180^{\circ}-\theta_{pp}^{\rm cm}$).
 Figure~\ref{fig5} shows the excitation energy distribution of the events
 for $\theta_{pp}^{\rm cm}$ from 0 to $7^{\circ}$ and
 $\phi_{pp}^{\rm cm}$  from  0 to $360^{\circ}$, summed over the beam energies 
 0.6, 0.7, and 0.8~GeV. 
 The shape of the spectrum is well reproduced ($\chi ^2/n.d.f.$=0.99)
 by the phase space distribution multiplied by the 
 Migdal--Watson factor describing the ${}^1S_0$ fsi \cite{watson} 
 including Coulomb effects.
 The event distribution over the angle between the relative
 momentum of the proton pair and its total momentum is nearly isotropic,
 but would allow a few percent of nonisotropic contamination
 to the differential cross section.
 The counting rates at high energies ($1.35$ and $1.9\,$ GeV)
 were rather low. 
 Therefore, in order to present the energy dependence of the process 
 for all measured beam energies, the three--fold cross section
 was integrated over the interval $0<E_{pp}<3$~ MeV and  
 averaged over the angular range $0<\theta^{\rm cm}_{pp}<8^{\circ}$,
 resulting in 
\begin{equation}
 \overline{\left(\frac{d\sigma}{d\Omega^{\rm cm}_{pp}}\right)}=
 \frac{N_{\rm cor}}{{L}^{\rm int}\cdot \Delta \Omega^{\rm cm}_{pp}}
 \cdot \frac{N_{\rm sig}}{N_{\rm sig}+N_{\rm bg}}\cdot f
\label{eq:1}
\end{equation}
 (Table~1). 
 Here $N_{\rm cor}=\sum_{i=1}^N 1/{(A_i\cdot \varepsilon_i)}$,
 $N$ is the number of selected proton
 pairs, $A_i$ and $\varepsilon_i$ correspond to 
 acceptance and detector efficiency  for registration of the  $i$--th pair. 
 The correction factor $f$, close to unity,
 accounts for several soft cuts applied during data processing.
 The acceptance was calculated as a function of 
 $E_{pp}$ and $\theta^{\rm cm}_{pp}$ assuming a uniform distribution in
 $\phi^{\rm cm}_{pp}$ and isotropy in the two proton system.
 The average detector efficiency was  $\varepsilon\approx 90\%$. \\
\par
\begin{figure}[htb]
\begin{center}
\epsfxsize=\columnwidth \epsfbox{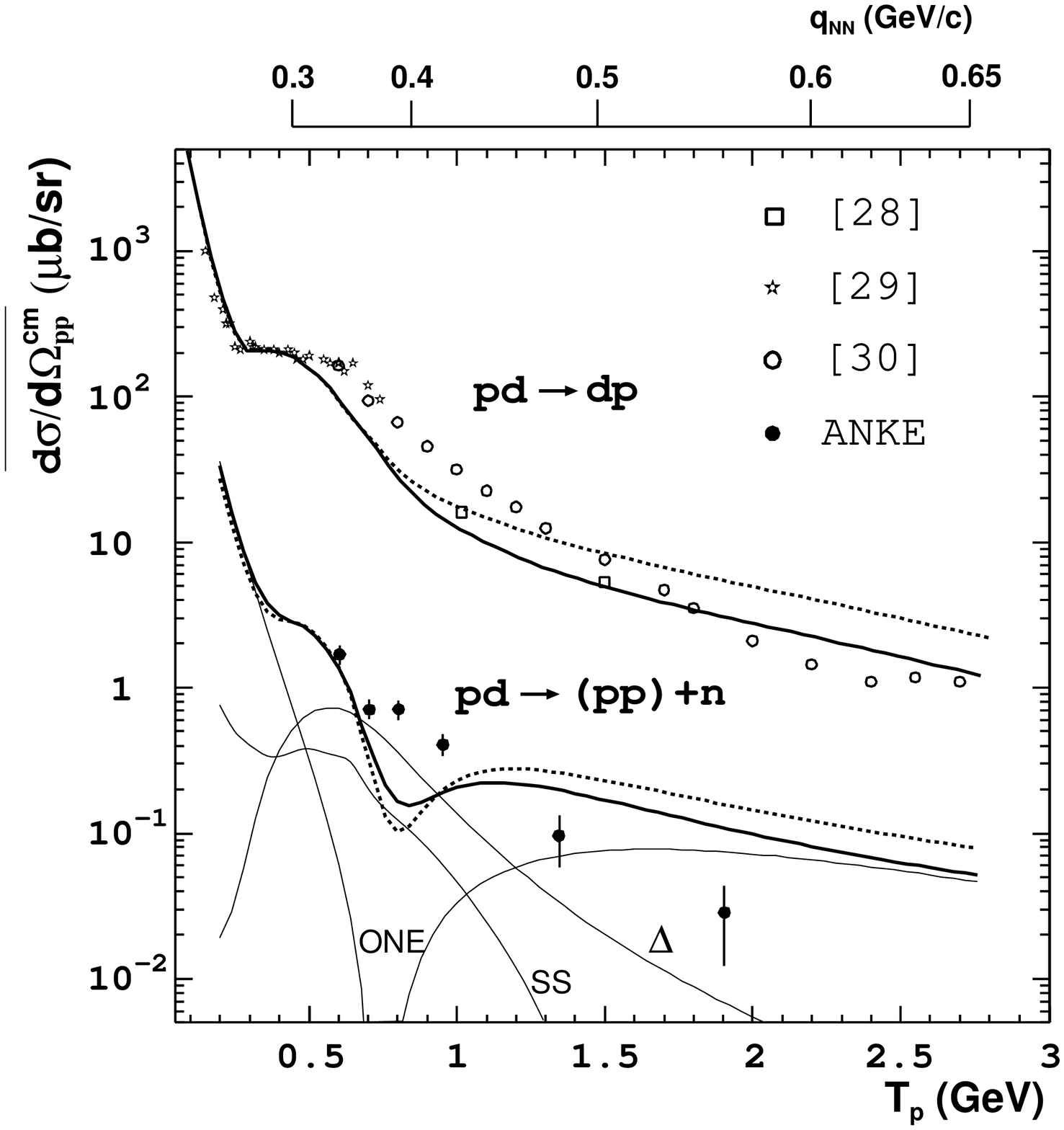}
\caption{\label{fig6} Measured cross section of the process                           
 $pd \to(pp)+n$ for $E_{pp} < 3$~MeV 
 versus proton--beam energy.
 The error bars include both statistical and systematic
 uncertainties (Table~1).
 Shown also are the $pd \to dp$ data 
 ($d\sigma/d\Omega^{\rm cm}_{\rm p}$)
 taken from Refs.~[28--30].
 The calculations with the ONE+SS+$\Delta$ model 
 are performed using the $NN$ potentials RSC (dotted line)
 and Paris (solid) \cite{uzikov2} (note also Ref. \cite{X}).
 The individual contributions of the ONE+SS+$\Delta$ model
 with the Paris potential are shown by thin full lines.
 The upper scale indicates the internal momentum of the
 nucleons inside the deuteron for ONE in collinear
 kinematics at $E_{pp}=3$~ MeV.}
\end{center}
\end{figure}

\begin{table}[htb]
\caption{Summary of the experimental results.
 $T_{p}$ denotes the beam energy, $L^{\rm int}$  the integrated luminosity, $N$ the number
 of events with $E_{pp}<3$~ MeV and pair emission angle
 $\theta^{\rm cm}_{pp}<8^{\circ}$, $N_{\rm cor}$ gives the number of events $N$,
 corrected for acceptance and detector efficiency, 
 $N_{\rm sig}/(N_{\rm sig}+N_{\rm bg})$ is the background correction,
 and $\overline{d\sigma/d\Omega^{\rm cm}_{pp}}$ denotes the cross section (see Eq.(1)).}
\begin{tabular}{|l|c|c|c|c|c|}
\hline
$T_{p}$ &$L^{\rm int}$ &$N$ & $N_{\rm cor}$
&$\frac{N_{\rm sig}}{N_{\rm sig}+N_{\rm bg}}$ 
&$\overline{d\sigma/d\Omega^{\rm cm}_{pp}}$ $\pm$ $\sigma^{\rm stat}$ $\pm$ $\sigma^{\rm syst}$  \\ 
$[$GeV]    &[cm$^{-2}\cdot 10^{34}]$ &              &     &    &[$\mu$b/sr]   \\
\hline
0.6   &1.41$\pm$0.12  & $339$  & 1403& $0.94\pm0.05$  &$1.72\pm 0.09\pm0.17$   \\
0.7   &1.93$\pm$0.17  & $227$  & ~872& $0.87\pm0.05$  &$0.72\pm 0.05\pm0.08$   \\
0.8   &2.38$\pm$0.20  & $305$  & 1050& $0.89\pm0.04$  &$0.72\pm 0.04\pm0.07$   \\
0.95  &1.28$\pm$0.11  & $112$  & ~337& $0.85\pm0.07$  &$0.41\pm 0.04\pm0.05$   \\
1.35  &0.69$\pm$0.06  & $~16$  & ~~45& $0.79\pm0.22$  &$0.10\pm 0.02\pm0.03$   \\
1.90  &0.74$\pm$0.07  & $~~9$  & ~~18& $0.62\pm0.27$  &$0.03\pm 0.01\pm0.01$   \\
\hline
\end{tabular}
\label{table1}
\end{table}
 The differential cross section 
 obtained as a function of beam energy is shown in Fig.~\ref{fig6}.  
 The energy dependence of the measured cross section
 is similar  to that of the 
 $pd\to dp$ process, but  its  absolute value is smaller 
 by  about two orders of magnitude. 
 There is no indication for the predicted dip in the  breakup cross section. 
 A comparison of the experimental results with the ONE+SS+$\Delta$ 
 calculations is shown also. At the lowest energies (0.6--0.7~GeV)
 the results for the Reid Soft Core (RSC) \cite{reid} and the Paris \cite{lacombe} 
 potential reproduce rather well the measured breakup cross section. 
 This energy range corresponds to the region where the $\Delta$(1232) 
 dominates in the $pd \to dp$ cross section.
 The theoretical curves for the breakup process 
 exhibit a shoulder at $\sim 0.5$~GeV as well.
 This indicates that in spite of the isospin suppression,  
 the contribution from the $\Delta$  is still important because of the
 nearby minimum of the ONE cross section.
 At higher energies, including  the region of the expected dip
 at 0.7--0.8~GeV, the model is in strong disagreement with the data.
 One should note that the ONE+SS+$\Delta$  model  underestimates
 the $pd \to dp$ cross section in the dip region 
 ($T_p\sim 0.8$~ GeV) as well \cite{X}.
 A possible explanation for this discrepancy is discussed 
 in Ref. \cite{uzikov}, where the contributions of 
 $NN^*$ components of the deuteron wave function are evaluated
 on the basis of a six quark model.
 Correspondingly for the breakup, effects from  $N^*$ exchanges and the 
 contribution of the $\Delta\Delta$ component of the deuteron can 
 possibly increase the cross section in this region and fill the dip.
 Other sizable  contributions may arise from intermediate states of  
 the $pp$ pair  at $E_{pp}>3$~ MeV, de--excited by 
 rescattering on the neutron in the final state.

\section{Conclusion}
 We report here the first measurement of the cross section of the
 $pd\to(pp)n$ reaction with a fast singlet $pp$ pair emitted in forward
 direction at beam energies between $0.6$ and $1.9$~ GeV.  The measurement
 was carried out in collinear kinematics close to those of $pd$ backward
 elastic scattering.  The known mechanisms of the $pd\to dp$ process
 describe reasonably well the measured breakup cross section at low
 energies (0.6--0.7~GeV).  At higher energies the calculations
 depend on the $NN$ interaction potential at short distances
 and disagree with the data.  
 Possible shortcomings of the
 model may be attributed at present to an inappropriate choice of the
 reaction dynamics or inadequate assumptions about the short--range
 structure of the deuteron. 
 The latter could be remedied by more detailed
 calculations using modern $NN$ potentials, which are in progress. \\
 We would like to emphasize that a study of the $pd\to (pp)n$ reaction
 with detection of $pp$ $^1{S_0}$ pairs provides a new tool to
 investigate the short--range $NN$ interaction.  
 For further insight, additional data, in particular polarization 
 measurements, are needed to provide a complete set of observables.
 These experiments are foreseen at ANKE.
 
\section{Acknowledgments} 
 We are grateful to J.~Haidenbauer (IKP, FZ J\"ulich) 
 for providing the scattering wave functions for the Paris potential.
 Valuable discussions with C.~Wilkin and his careful reading of 
 the manuscript are appreciated.
 We would also like to acknowledge in particular the early contributions
 by O.W.B.~Schult. Some of us acknowledge the warm hospitality
 and support by FZ J\"ulich. This work was supported by the BMBF WTZ 
 grants KAZ 99/001, RUS 00/211, and RUS 01/691, and by the 
 Heisenberg--Landau program.


\begin{thebibliography}{00}
\bibitem{X} According to  Ref.~[3] 
 a coefficient of 0.8 in the ONE amplitude arising from the distortion
 of plane waves in the initial and final states is used to match the 
 absolute value of the $pd \to dp$ cross section at $T_{p} <0.3$ GeV.
\end{thebibliography}
\end{document}